# Langar: An Approach to Evaluate Reo Programming Language

Mohammad Reza Besharati[1]
Mohammad Izadi[2]
1- besharati@ce.sharif.edu, PhD Candidate, Sharif University of Technology, Tehran, Iran, Corresponding Author.
2- izadi@sharif.edu, Assistant Professor, Sharif University of Technology, Tehran, Iran.

## Abstract

Reo is a formal coordination language. In order to assess and evaluate its capabilities, we need a multi-perspective Language Evaluation Framework. Langar (**Lang**uage **A**nalysis for **R**eo) is a framework aimed to provide such an evaluation method. In this paper, we introduce Langar. Based on a review on various language evaluation methods, a tool-kit for useful evaluation techniques are provided. After Reo Evaluation, this method and tool-kit also could be used for another programming, computational and even natural languages. Furthermore, two suggestions for some future efforts and directions are provided for software engineering and software methodology communities.

## 1. Introduction

Programming languages can be a common phenomenon in the following areas 1) mathematics, logic, and computing sciences, 2) semantics, understanding, and the human mind, and 3) engineering, industrial, operational, and servicing applications.

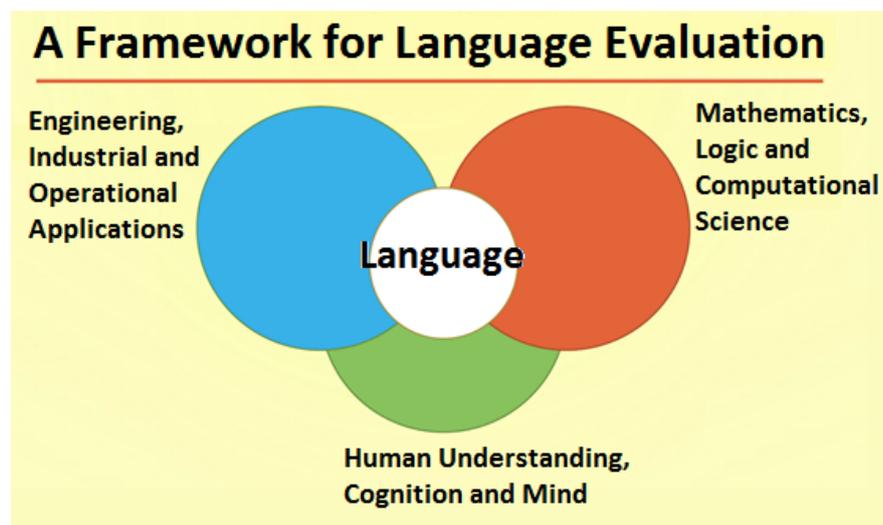

*Figure 1- Programming languages, a common phenomenon in the three prominent areas, in Langar Framework*

For each of the above areas, we can consider a point of view (or points of view) and a criterion (or criteria) for evaluating and analyzing a given language. Some points of view or criteria have been conventionally important and reflected in the literature review of programming languages. Table 1 presents some of the most important criteria and points of view for each of the above areas.

Given the availability of various programming languages (e.g., C and Reo), programming tools (such as compilers and MDD tools), software development systems (e.g., infrastructures and platforms), and ultimately, software evolution ecosystems, such criteria and points of view seem to be essential for scientific, technical, quantitative, and qualitative evaluations and comparisons.

*Table 1- Common points of view and criteria to evaluate programming languages for each area*

| Domain of Mathematics, logic, and computing sciences | Domain of Semantics, understanding, and the human mind | Domain of Engineering, industrial, operational, and servicing applications |
|---|---|---|
| Computing expressive power | Expression and specification | Applicability and usefulness |
| Complexity | Legibility for humans and the final user | Up scaling |
| Underling logic and model | Composition and construction styles | Affordability |
| Evaluation, confirmation, and testing | Integration process | Paradigms and the Engineering Process |
| Supporting modalities | Analytical and synthetic point of view | Language and application engineering |
| Evaluation, soundness, and completeness | Supporting the development, evaluation, and calculation | Scalability and adaptability |
| --- | --- | --- |

## 2. Evaluation; Various Approaches

There are various methods and approaches for evaluating systems and software attributes. Some have moved further and provided systematic or non-systematic frameworks for evaluating systems, methods, and tools (e.g., programming languages) in the field of software engineering. In summary, the following are the most important approaches to evaluate systems, attributes, and software languages.

1. Evaluation based on the definition and confirmation of features
2. Evaluation based on factor analysis
3. Evaluation based on experimentation and benchmarking
4. Evaluation based on case studies and summarizing real experiences
5. Evaluation based on field studies and the literature review
6. Evaluation based on technology futures studies
7. Evaluation based on the opinions of experts, users, and practitioners (developers)
8. Criteria-based analysis and evaluation
9. Evaluation based on modeling perspective
10. etc.

In the following, each of the above approaches and their backgrounds is discussed.

## 2.1. Evaluation based on the definition and proving of features

This approach, according to a kind of mathematical and formal theory (or theories) and/or a paradigm-based and informal theory (or theories), defines specific features of the language (or programs and systems that are written by that programming language). Then, using mathematical tools or other methods of reasoning (relevant to the Paradigms of the Context, which is a mathematical Paradigm or a non-mathematical Paradigm), we prove the characteristics of the language. A well-established example of this evaluation approach is the Turing completeness proof for programming languages by referring to the theory of languages and machines and the theoretical model of the Turing machine and establishing a formal and mathematical correspondence between the computational model of language and the computational model of the Turing machine. Another example is the use of the Lambda calculus to define and prove a variety of features for programming languages [1]. The critical point in such approaches is the existence of a basic theory or computational model (e.g., Turing machine, Lambda calculus, rewriting logic, etc.) to propose a framework for defining and proving features.

The approach of defining and proving features to evaluate programming languages can be classified into two following categories.
1. Evaluation based on mathematical Paradigm and formal methods (Figure 2)
2. Evaluation based on other paradigms (Figure 3)

$$E((\lambda_n x.e)e') \longrightarrow E(e[x/e'])$$
$$E((\lambda_v x.e)v) \longrightarrow E(e[x/v])$$

**E1** $\varphi(e)$ is an $\mathcal{L}'$-program for all $\mathcal{L}$-programs $e$;

**E2** $\varphi(F(e_1, \ldots, e_a)) = F(\varphi(e_1), \ldots, \varphi(e_a))$ for all facilities $F$ of $\mathcal{L}'$, i.e., $\varphi$ is homomorphic in all constructs of $\mathcal{L}'$; and

**E3** $eval_{\mathcal{L}}(e)$ holds if and only if $eval_{\mathcal{L}'}(\varphi(e))$ holds for all $\mathcal{L}$-programs $e$.

*Figure 2- An example of defining mathematical and official features for programming languages. In this example, the Lambda calculus has been used as a framework and platform for defining and proving features [1].*

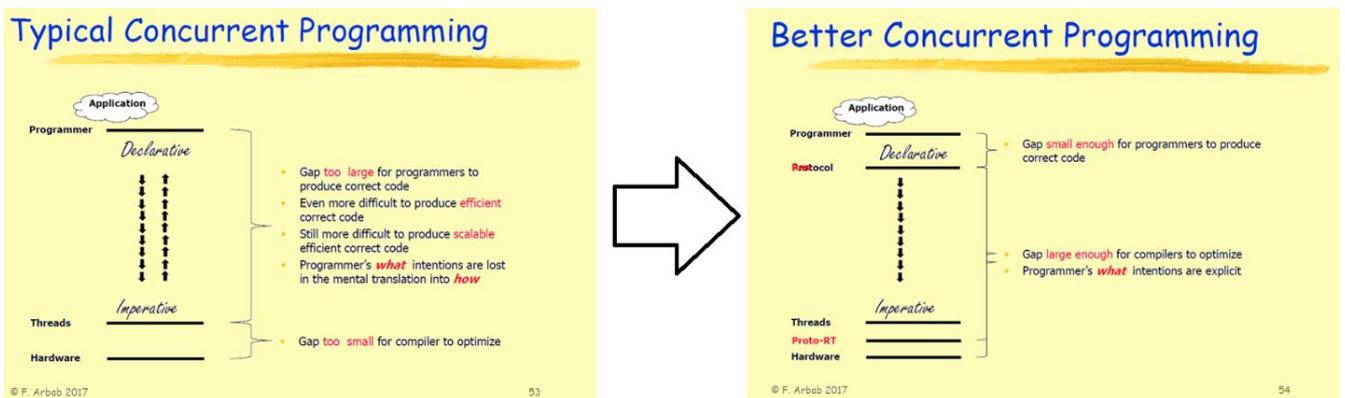

*Figure 3- An example of defining features and evaluating them in a Paradigm rather than mathematics. Here, concepts and principles related to the programming (in particular, the concurrent programming language) have been considered as a base for definitions and reasoning [10].*

Mathematical discourse-based evaluation is a well-known approach. It is standing for a large portion of the research literature in computer science. The second case (i.e., evaluation based on other discourses), however, has been less welcomed in research in this field for various reasons. It may be due to the existence of precise theoretical tools and ready-to-use infrastructure for the first approach (i.e., mathematical evaluation) that has led researchers to assume that "evaluation with a paradigm other than formal and mathematical paradigms will be general-speaking and does not enable us to make an accurate evaluation and correct proof."

However, this approach has been challenged in recent years in many fields of cybernetics and engineering, such as system engineering and software engineering. Today, we are seeing the appearance of approaches that are not accurate and mathematical (i.e., in their algebraic, logical, computational, and classical senses), but are systematic, coherent, and reasoning. Therefore, they can be thought to carry a kind of mathematics and non-classical calculation. Examples are as follows:

- Types of inaccurate but systematic modeling approaches of systems and organizations (e.g., UML, KAOS, CMMI, GQM, AHP, BBN, methods to model concepts in dimensional spaces, techniques of modeling concepts in the form of components of an architecture, approaches to modeling concepts in the form of a network, structural analysis, behavioral analysis, causal analysis, analysis of generators and grammars, network analysis, etc.)
- And/or intelligent evaluation techniques based on artificial intelligence and soft computing (e.g., statistical methods, techniques based on linear algebra modeling, fuzzy methods, probabilistic methods, word computations, neural networks, cognitive computations, association, induction, inference, analogy, etc.)
- And/or are phenomenological approaches (e.g., studies of large-scale networks, complex and similar systems and such as graphic pattern recognition, graphical transformations, motif detection, measurement of quantitative-statistical features of large-scale networks, etc.)
- And/or anthropological, sociological, discourse-oriented, and futuristic approaches that are seemingly quite far from classical mathematics but very close to some kind of reasoning and adherence to some kind of systematic theoretical framework (such as HCI, ethnographic studies on programmers, companies, and organizations working in the software industry, usability studies, organizational studies, demand engineering, etc.)

A paradigm can be defined in a variety of ways and approaches, i.e., a theoretical framework as a basis for reasoning that holds definitions, rules, and principles, and is governed by a kind of mathematics and a kind of logic, even if it is basic and simple. In some computer science articles and research, this concept is referred to as a theory or a framework (rather than discourse). Nevertheless, apart from the name we allocate (i.e., discourse, theory, framework, method, model, etc.), the central topic does not make a difference to our research objective. If we have no mathematics and require it, we can construct it in the form of a system of reasoning and discourse, even rudimentarily. And by which, we can define and prove the characteristics of a language.

## 2.2. Evaluation based on factor analysis

According to the analytical approach (decomposing a whole into components and then reconstructing the whole from the components, which are called respectively analysis and synthesis), it is feasible to evaluate the various features of a soft system or a language. Figure 4 gives an example of an analytical-causal structure for a set of factors influencing the overall quality of software. In this manner, an unexplained concept of "software quality" is defined in the form of an analytical structure, in a systematic, somewhat more tangible, and quantitative way. A similar hierarchical and analytical structure can be extracted for the factors involved in the various features of a programming language.

Bayesian and probabilities coefficients, statistical distributions, and weights learned by a learning feedback mechanism (such as neural networks) can be placed on the edges or nodes of an analytical structure. By this operation, the analytical structure can be transformed from a purely descriptive model to a computational and operational model for measuring the values of the properties under evaluation.

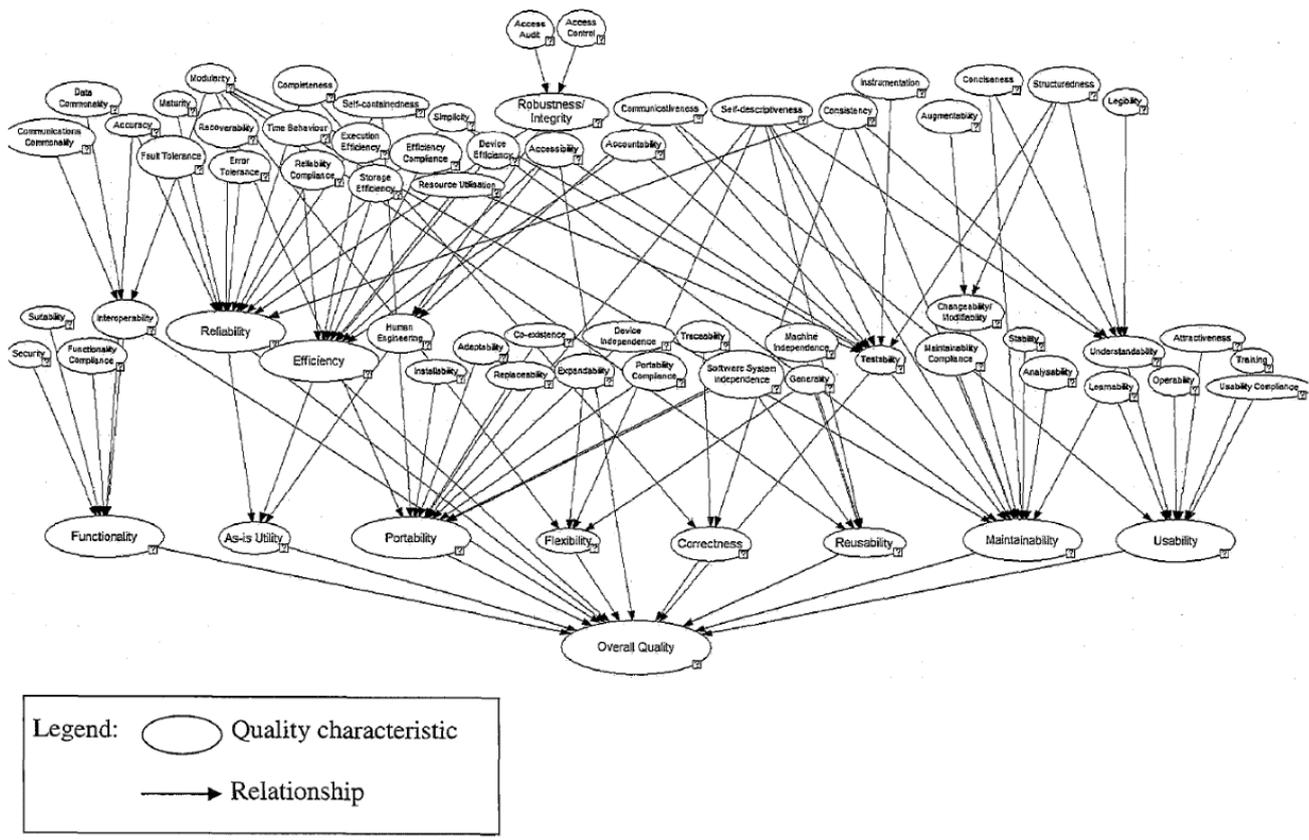

*Figure 4- An analytical-casual structure for a hierarchy of factors influencing the overall quality of software [5].*

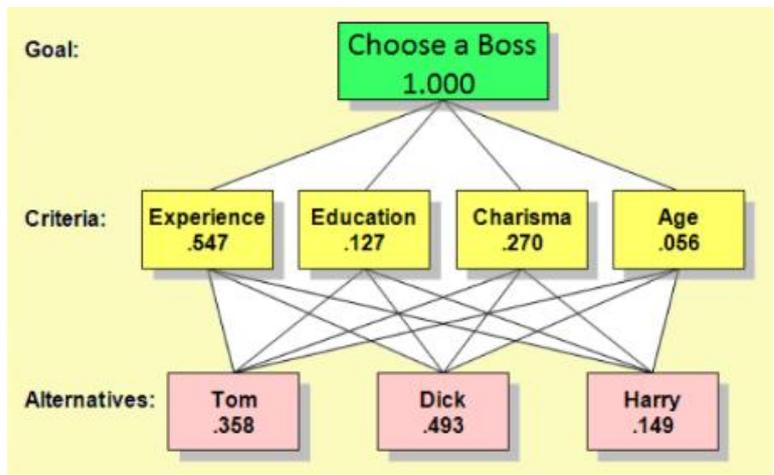

*Figure 5- An example of the AHP method for implementation of evaluation calculations using an analytical structure [4].*

**Suggestion 1.**

It is possible to extract a hierarchical analytical structure for the properties of programming languages and to learn the coefficients on its edges by applying machine learning algorithms by inputting the analysis of dozens of different classical languages into one, The "evaluation ruler" could be constructed to evaluate emerging and newer languages (such as Reo).

## 2.3. Evaluation based on experimentation and benchmarking

In this approach, the various features of soft systems and languages can be evaluated by defining tests, experiments, standard evaluation procedures, benchmarks, platforms, and so on. A well-known application of this approach is to evaluate the features related to the efficiency and performance of the systems. For example, for the Reo language itself, this evaluation approach has been used to compare the performance of code derived from classical concurrency programming with code derived from compiling Reo circuits.

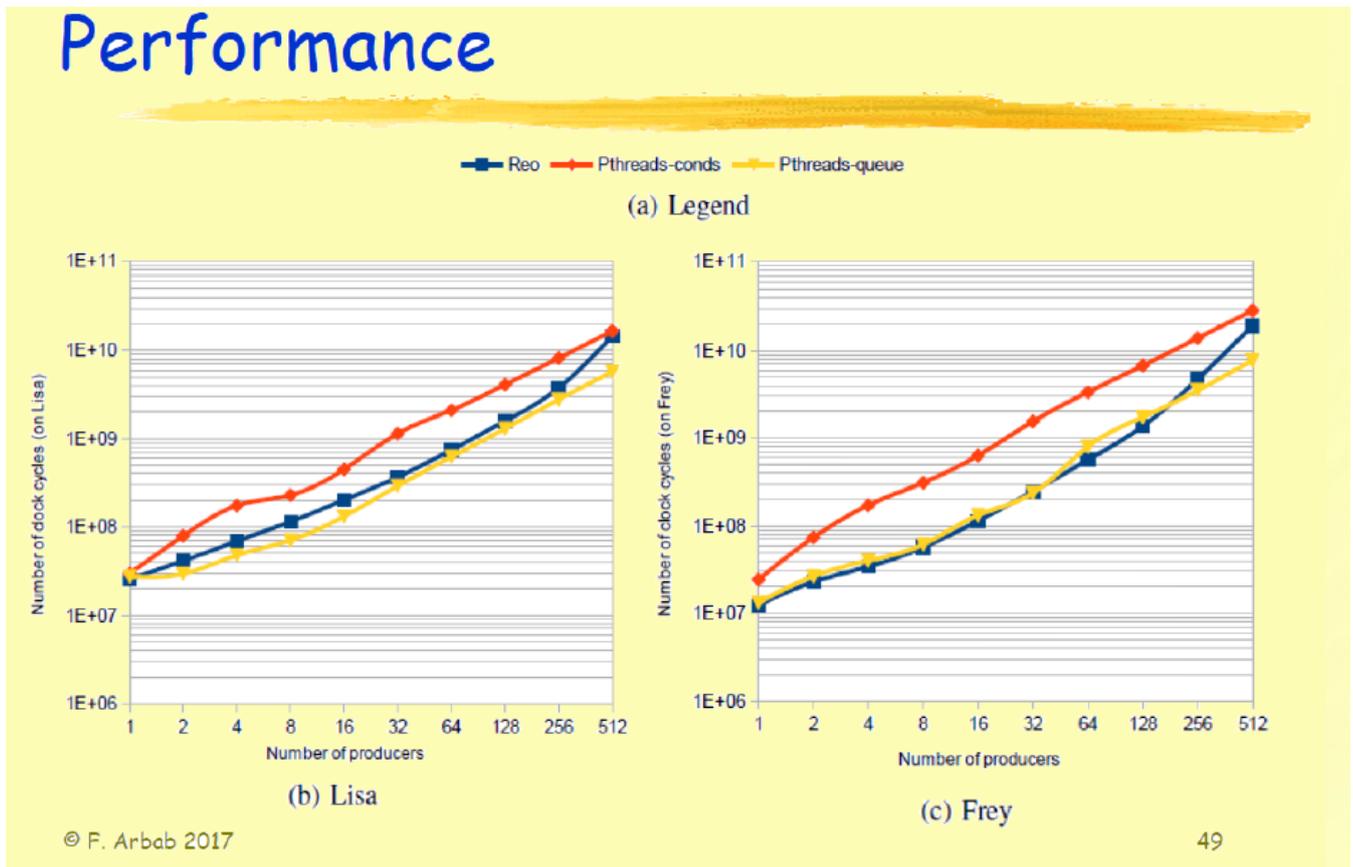

*Figure 6- Evaluation of the performance by experimentation and benchmarking to compare the performance of code derived from classical concurrency programming with code derived from the Reo programming language[10].*

## 2.4. Evaluation based on case studies and summarizing real experiences

In this approach, evaluation is performed through the implementation of a case or cases of applications and uses of a tool, approach, solution, method, or language. This case study may be on a particular example and details of a well-known problem, or to define an issue or issues specifically for the case study of a soft system or language. The application or use maybe not solely for the purpose of evaluation, and we may see a case or cases of the real experience of the use and application of that soft system or language, or data and documentation of real industrial experiences and non-industrial use of a language after its application. In this case, by summarizing real experiences (stories on failure, success stories, reports and analyzes of success and failure, post-mortem analysis and project summary documents, technical reports, case studies, Practical application reports, reports of industrial and real-world applications, etc.), we can evaluate based on data obtained from the practical reality and application of the system in the environment and real problems.

This type of evaluation has been used in the Reo language and can be utilized for more evaluation of language and Reo approaches. Figure 7 shows an example of Reo's application in designing a real-world application system.

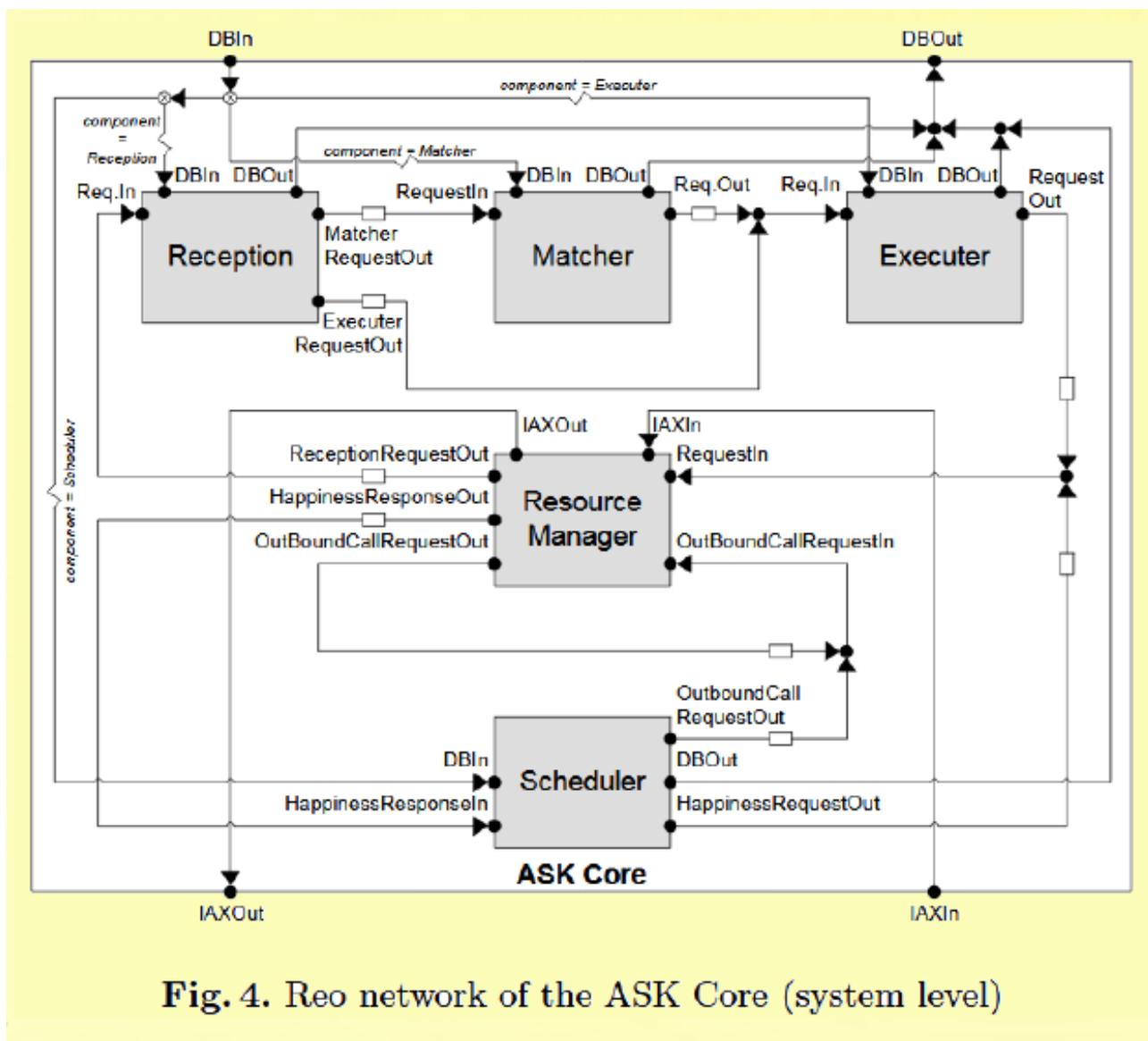

Figure 7 - A Reo circuit related to one of the industrial and real-world applications of Reo [3].

**2.5. Evaluation based on field studies and research literature**

A comprehensive evaluation can be carried out by referring to studies conducted over the years for a programming tool, method, or language by various researchers. Evaluation approaches such as systematic domain research, systematic mapping studies, and critical studies fall into this category. Importantly, these approaches rely on the massive research efforts of researchers and teams that have had different experiences with that language or programming tool with different approaches.

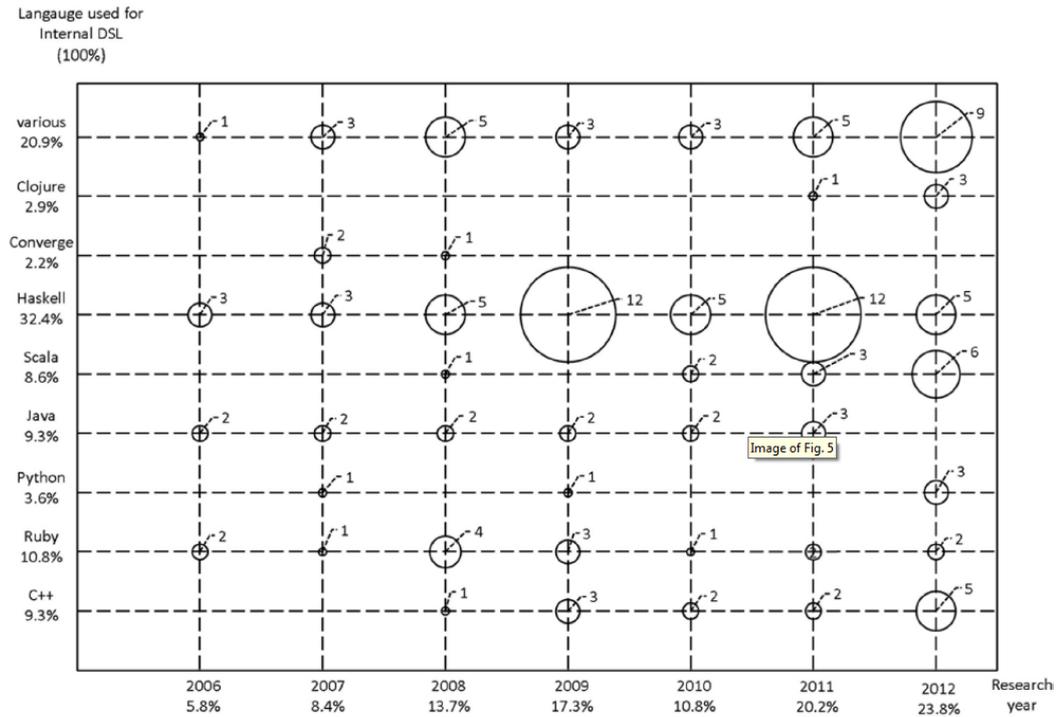

*Figure 8 - Visualization of a systematic map in the form of a bubble plot [6].*

## 2.6. Evaluation based on technology studies

Considering that programming languages and tools are a component of technological solutions in the field of information technology (IT), this dimension or point of view is very important and useful for their analyses. There are different methods and approaches for technology studies and, indeed, it is a distinct scientific, research, and specialized field of study. Approaches to evaluating through technology studies include "application studies", "technology and eco-bio technology analysis, "technology future studies and analysis", "innovation, creativity and interdisciplinary studies", "evaluation based on industrial, commercial and organizational standards, requirements and procedures "and" evaluation based on adherence to rules and regulations".

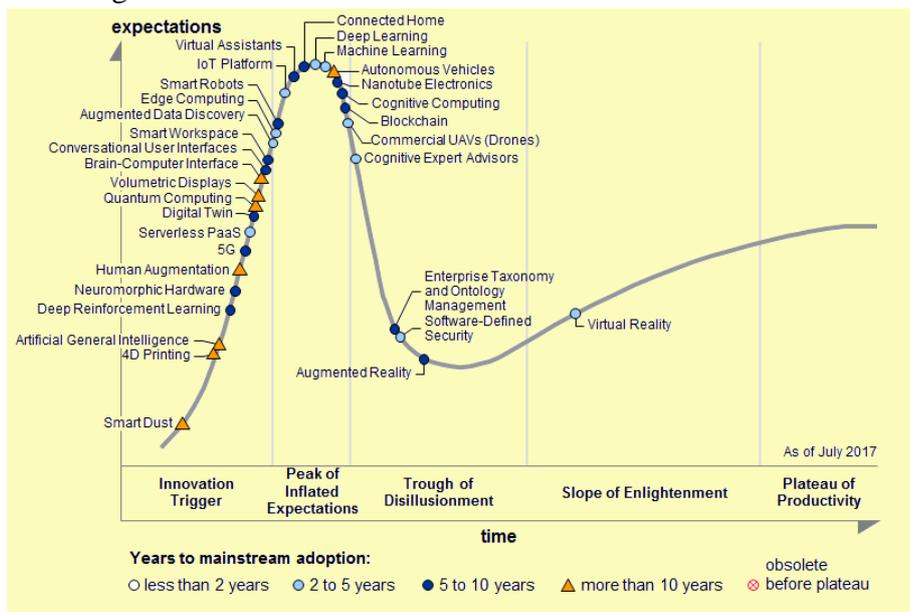

*Figure 9- The Gartner Hype Cycle graph for technologies emerged in 2017. This graph, which is based on a research approach in technology studies with the same name, is used for analyzing the status of technology in terms of being in the various phases of growth.*

## 2.7. Evaluation based on the opinions of experts, users, and practitioners (developers)

An evaluation view of tools, languages, and environments can be achieved by utilizing and summarizing the opinions of experts, users, and practitioners (developers). There are approaches for collecting and then systematically summarizing the opinions of experts. For example, in [7], a summary of the opinions of experts in a conference on "A Theory for Software Engineering" is presented, with an interesting technique and format used to summarize the opinions. For example, conceptual architecture has been derived from opinions.

A well-known approach for summarizing opinions and extracting coherent theories from the opinions of experts is "Grounding Theory", which has also been used in software engineering [8].

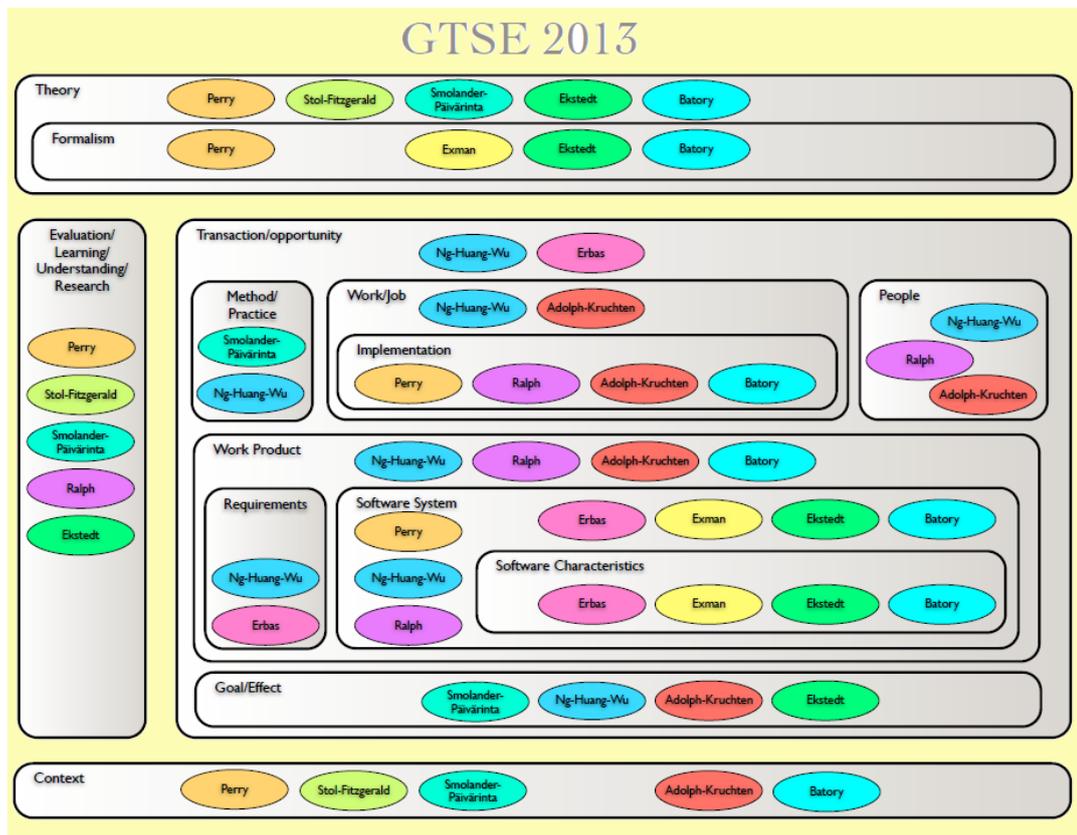

*Figure 10 - A conceptual architecture extracted from the summary of opinions of experts on "A Theory for Software Engineering" [7].*

## 2.8. criteria-based evaluation and analysis

In this approach, a set of criteria is first defined or extracted for evaluation. Then, based on these criteria, the system, method, approach, language, or tool is evaluated [9]. This approach is one of the most widely used evaluation techniques in the field of software engineering.

## 2.9. Evaluation based on modeling perspective

Real engineering is feasible without modeling. Models enable us to build computational, accurate, and calculated engineering artifacts. The elements of a general metamodel for the modeling problem include the system under study (SUS), the modeling point of view, the modeler, and the modeling language. Therefore, software engineering methods, tools, languages, and approaches can be evaluated based on these four basic components, which are inherent in the modeling problem.
- For what SUS does this language is suitable?
- What modeling points of view does the model support?

- Which modelers can do easily utilize this tool? And
- What is the mode of expression, symbolism, semiotics, semantics, structure, and construction of the modeling language?

In this area, the modeling perspective can be considered as one of the important aspects of evaluation. The classic and basic modeling perspectives in system engineering and software engineering include attributes, functions, structural and static perspectives (e.g., the main UML Class diagram), behavioral and dynamics perspectives (e.g., the main UML activity diagram), the generative perspective (e.g., generative grammars), the causal perspective (e.g., the cause and effect relationships diagram), the evolutionary perspective (e.g., the evolutionary appearance), and so on. The problem can be viewed from any perspective. The number of perspectives can be infinite in considering composite, multidimensional, and specific perspectives of the problem and scope.

One of the important perspectives for modern, complex, complex adaptive, and multi-agent systems is the perspective of "agents and coordination and interaction between them", which can be termed the coordination perspective. For complex systems (including human, organizational, physical, cyber, software, etc.) the protocol and the logic of interaction and coordination between the autonomous, orthogonal, interacting, and concurrent factors that play a role in the system should be somehow modeled. Furthermore, the logic governing the affairs and complexities of the system should be expressed globally and intersectionally.

> **Suggestion 2.**
> It seems that there is no efficient, flexible, and appropriate support on the current and prevalent approaches of software engineering (and even to some extent system engineering) in terms of "coordination" modeling and "interaction protocol modeling". For example, in UML diagrams for software engineering, there is no model or diagram specific to the perspective of coordination and protocol of interaction between agents, which is a leading shortcoming for the application of modern methods for complex systems in the field of software development, evolution, analysis, and design. The Reo language and Reo-based techniques are recommended to be introduced as a remedy for this gap in the field of modeling techniques and languages in software development, analysis, and design.